\documentclass[fleqn,usenatbib]{mnras}

\usepackage{newtxtext}

\usepackage[T1]{fontenc}

\RequirePackage{fix-cm}

\usepackage{graphicx}	
\usepackage{amsmath}	
\usepackage{newpxmath}
\usepackage{orcidlink}

\title[Short title, max. 45 characters] {Variable ADAF disk as the origin of Changing-Look AGN}

\author[Chun Xu]{Chun Xu$^{1}$\thanks{E-mail: chun.xuu@shao.ac.cn}\orcidlink{0009-0009-2507-5977}
\\
$^{1}$Shanghai Astronomical Observatory, Chinese Academy of Sciences, Shanghai 200030, China}

\date{Accepted XXX. Received YYY; in original form ZZZ}

\pubyear{\the\year{}}

\begin{document}
\label{firstpage}
\pagerange{\pageref{firstpage}--\pageref{lastpage}}
\maketitle

\begin{abstract}
 We propose that changing-look AGN transitions arise from variations in the size of the inner ADAF disk. The AGN accretion disk consists of an outer thin disk and an inner thick ADAF component, whose size is intrinsically unstable and evolves over time. The size variations of the ADAF are governed by a parameter $\eta$, which represents the turbulence strength within the accretion flow. $\eta$ also determines the accretion rate onto the central black hole and controls jet formation and outflow rate, with the latter regulating the line-of-sight absorption. From the perspective of a variable ADAF, changing-state and changing-observation AGN are two sides of the same coin. We further discuss gigahertz-peaked and compact steep-spectrum radio sources as possible manifestations of intermediate-to-large scale ADAFs. Finally, we propose that AGN unification models should include both orientation and ADAF size as key parameters.
\end{abstract}




\section{Introduction}

Changing-look active galactic nuclei (CL-AGNs) are a remarkable class of 
extragalactic sources that exhibit dramatic spectral transitions between 
different AGN types on timescales of months to decades. First identified 
in the 1970s with the discovery of Mrk~1018's transition from Seyfert 
1.9 to type~1 \citep{Cohen1986}, these systems challenge the classical 
unification model by demonstrating that AGN classification is not solely 
determined by orientation. The transitions manifest as the appearance or 
disappearance of broad emission lines---most notably Balmer lines---which 
respond to changes in the accretion rate and inner disk geometry 
\citep{Ricci2023}. Two primary mechanisms are invoked to explain these 
phenomena: intrinsic changes in the accretion state, where the inner 
disk transitions between a standard thin disk and a radiatively inefficient 
flow \citep{Veronese2024}, and variable obscuration by clumpy material in 
the broad line region or torus \citep{Risaliti2002}. The most compelling case for intrinsic 
transitions is 1ES~1927+654, which transformed from a ``naked'' Seyfert~2 
to a broad-line AGN within months in 2017--2018, with newly formed broad 
line clouds detected on eccentric orbits \citep{Trakhtenbrot2019}. 
Recent systematic surveys have identified hundreds of CL-AGNs, revealing 
that such transitions may occur in 10--20\% of AGN over cosmic time 
\citep{Birmingham2025}, fundamentally altering our understanding 
of black hole growth and AGN duty cycles.

From a theoretical perspective, \citet{Xu2026a} extended the advection-dominated accretion flow (ADAF; \citealt{Narayan1994, Narayan1995}) to a more general framework. By assuming that the binding energy released in the disk can be temporarily stored in turbulence, an ADAF disk can form without requiring the two-temperature assumption for ions and electrons. This newly developed ADAF disk is dense, geometrically thick, and optically thick, making it applicable to diverse accretion systems including AGNs, X-ray binaries, and young stellar objects. Central funnels are established within the thick ADAF disk, where jets may form and become collimated. In a subsequent paper, \citet{Xu2026b} proposed that the ADAF disk is intrinsically unstable; its radial extent may vary due to instabilities. This variable ADAF disk can explain numerous observed behaviors in black hole X-ray binaries, including: (i) the X-ray burst light curves and the q-shaped tracks in hardness--intensity diagrams, both related to changes in ADAF size; (ii) the frequencies of quasi-periodic oscillations, tied to the ADAF scale; (iii) the hard X-ray power-law spectrum, connected to the Kolmogorov spectrum of turbulence; and (iv) variations in accretion rate, linked to ADAF size. 

In this work, we propose that the variability of the ADAF disk can also account for many observed phenomena in changing-look AGNs. We describe our model in Section~2. Then we compare different individual CL-AGNs with the model in Section~3. In Section~4 we discuss the peaked-spectrum/gigahertz peaked-spectrum (PS/GPS) and compact steep-spectrum (CSS) radio sources within the variable ADAF model. Summary and discussions are presented in Section~5.

\section{Model description}

The accretion disk around the central black hole consists of an outer thin disk \citep{Shakura1973} and an inner thick ADAF disk \citep{Xu2026a,Xu2026b}. The inner ADAF disk is unstable, and its size may vary. In extreme cases, it may disappear and be replaced by an extension of the outer thin disk.

The geometry of the inner thick ADAF and outer thin disk can be described following \citet{Xu2026a,Xu2026b}. We first compute the equipotential contours \citep{Xu2026a,Frank2002} at various radii using the corresponding $\eta$ values, where $\eta$ is a parameter that represents the turbulent energy in the disk in units of the local Keplerian energy ($\eta$ ranges between 0 and 1). For illustrative purposes (not as a real physical relation between $\eta$ and $R$), we assume that $\eta$ follows a sigmoid function of radius $R$, with $\eta \sim 0$ at large radii and $\eta \leq 1$ as $R\sim0$ (or as $R$ goes to black hole). We then define the disk by drawing an envelope over these contours. Fig.~1 shows, for comparison, the cross sections of two ADAF disks indicated by red envelopes, where the maximum $\eta$ is 0.46 on the left side and 0.92 on the right side, respectively. The maximum values are reached in the innermost region of the disk ($R \sim 0$). Both the left and right ADAFs are plotted using the same sigmoid function, differing only in their maximum $\eta$ values. Fig.~1 clearly shows that the funnel opening angle is larger for smaller $\eta$ (left panel). According to \citet{Frank2002} in §10, the half opening angle $\theta$ of the funnel is determined by the equation:
$\sin^2(\theta) = -e =1-\eta$, where $e$ is the specific total energy of the disk fluid and is related to $\eta$ by $\eta=e+1$. The total specific energy of the jet/outflow is $e_j=1+2\eta$, and the kinetic energy of the jet/outflow after subtracting the escape energy (or gravitational energy) is $e_{\rm kj}=1+2\eta -2 = 2\eta -1$ \citep{Xu2026a}. Note that $\eta \sim 0.5$ represents the minimum value required for jet/outflow formation.

\begin{figure}
 \includegraphics[width=\columnwidth]{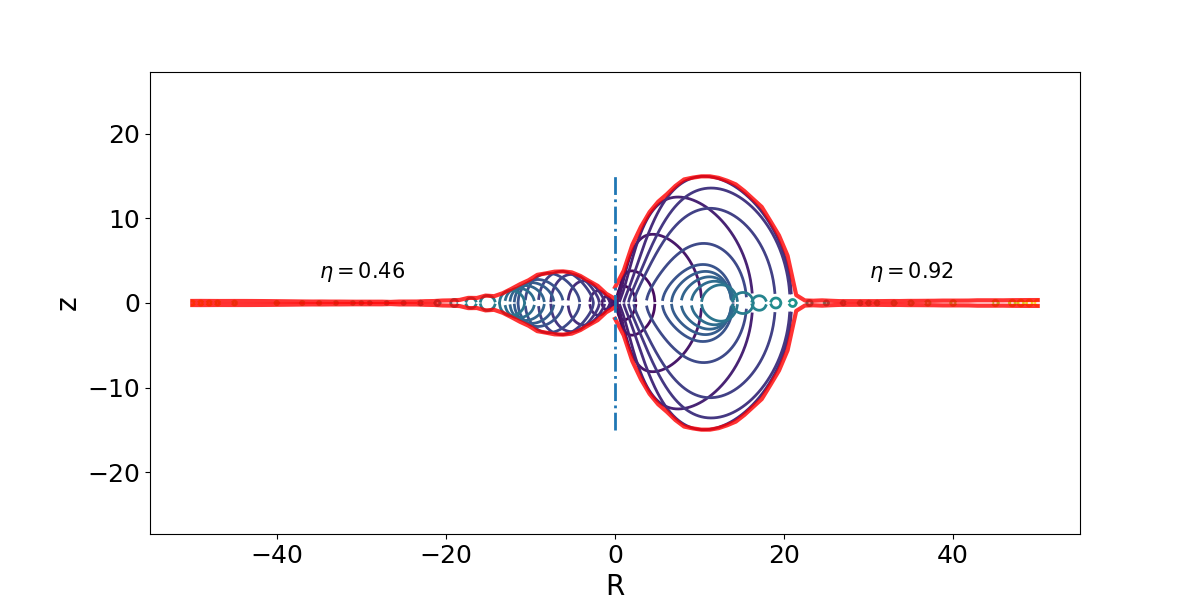}
 \caption{The geometry of the inner ADAF thick disk, which connects with the outer thin disk, is depicted by the red envelope. Two ADAFs are shown on the left and right for comparison. The dot-dashed line marks the separation at $R=0$. The bluish thin lines represent equipotential contours at different radii, each with varying $\eta(R)$ values. The left side of the plot displays an ADAF with a maximum $\eta = 0.46$, while the right side shows an ADAF with a maximum $\eta = 0.92$. The coordinates are on arbitrary scales.}
 \label{fig:ADAF_disk4}
\end{figure}

The emission from AGN originates in several distinct regions. X-rays and possibly $\gamma$-rays are produced in a corona surrounding the central black hole. In radio-quiet AGN, the size of the corona is approximately $10~R_{\rm g}$, where $R_{\rm g}=GM/c^2$ for a black hole of mass $M$ \citep{Laha2025}. In radio-loud AGN, the jet base may also contribute to the X-ray emission, and the effective size (or length of the "corona") can extend up to $130~R_{\rm g}$ \citep{Dogruel2020}. The X-ray spectra are likely associated with synchrotron emission, inverse Compton scattering \citep{Laha2025}, or with a power-law distribution of energetic particles induced by Kolmogorov turbulence in an ADAF torus \citep{Xu2026b, schlickeiser2002}.
In fact, we tend to believe that both the corona itself and the power-law spectra arise more likely from turbulent blobs created in the funnel, as discussed in the next chapter on the source 1ES 1927+654. The iron $K\alpha$ line is a reflection component from the accretion disk \citep{Reeves2006} or the inner edge of the ADAF torus.

The Big Blue Bump (BBB) represents the thermal emission from the accretion disk, extending from the optical to the soft X-rays \citep{Shields1978, Malkan1982, Elvis1994}. Modern interpretations favor warm Comptonization as the origin of the soft X-ray tail of the BBB \citep{Kubota2018}. The BBB is considered the primary UV energy source for the broad emission lines. Radio emission mostly originates from the jet in radio-loud or intermediate radio AGNs.

The broad emission lines in active galactic nuclei originate from photoionized gas located in the broad line region (BLR), which extends over distances of approximately 10 to 1000 $R_{\rm g}$ \citep{Peterson1993, Kaspi2000, Bentz2013}. In this region, the high-velocity Keplerian motion around the central supermassive black hole induces Doppler broadening of recombination and collisionally excited lines, such as H$\alpha$ and H$\beta$. The BLR is thought to reside in several types of environments: one consists of discrete blobs scattered around the central black hole \citep{Blandford1982,Peterson1993}, and the other corresponds to the inner region of a thin disk \citep{Livio1997, Eracleous2003}. As will be discussed later, the blobs in the BLR are primarily supplied by outflows from the central funnel, or some from ADAF disk itself.

The size of the ADAF disk is variable, but we can still obtain a rough estimate of it under different conditions. The inner radius of the ADAF is approximately around the innermost stable circular orbit (ISCO), i.e., about $6R_{\rm g}$ or even smaller. This is supported by observations \citep{Lu2023,Reeves2006} and is also a requirement for the formation of radio jets \citep{Xu2026a}. The outer radius of the ADAF is inferred from the truncated thin disks in NGC 7469 and Mrk 352, at $\sim$ 35–125 $R_{\rm g}$  and 50–135 $R_{\rm g}$ \citep{Kumar2023}, respectively, or from the black hole XRB GX 339-4 \citep{chainakun2021}, whose truncated thin disk varies between $10$ and $55 R_{\rm g}$ during the source luminosity decreasing phase. The truncated thin disk essentially marks the outer radius of the ADAF (Fig.~1). We may adopt $100 R_{\rm g}$ as a typical value for the outer boundary of the ADAF in AGN. As seen in Fig.~1, the height of the ADAF is approximately the same as its outer radius (when $\eta \sim 1$ ). The radius of the ADAF differs among different types of AGNs (to be discussed in the next chapter) and may vary over time. The size of the ADAF is positively nonlinearly related to the parameter $\eta$ \citep{Xu2026b}.

Let us examine how different types of AGNs are related to different ADAF disk sizes (or $\eta$). When $\eta$ is small, say below 0.2 (just an example, not a calculation), the ADAF is small and a thin disk extends deep in. In this case, both the BBB and broad lines should appear. However, no jets or outflows form from the central funnel. Some outflows may emerge (from the disk) \citep{Torrespapaqui2020}. This typically corresponds to a Seyfert 1 AGN.

For $\eta$ between 0.2 and 0.5, the ADAF expands and occupies part of the thin disk region. As a result, we may expect the big blue bump (BBB) to decrease, with the EUV decreasing particularly significantly. Consequently, the broad emission lines will also weaken. In this case, the AGN would more closely resemble a Seyfert~2 galaxy, or perhaps a mixed state between Seyfert~1 and Seyfert~2.

For $\eta$ between 0.5 and 0.7, the ADAF expands further, covering a larger area of the thin disk and further reducing the BBB UV emission. The half-opening angle of the funnel becomes smaller; $\sin^2\theta = 1 - \eta$ gives $\theta$ between $45^\circ$ and $33^\circ$. The blob energy from the funnel is $2\eta - 1$, ranging between 0 and 0.4 (Keplerian energy). Therefore, we expect strong outflows emerging from the funnel and extending outward in a conical shape. The X-rays from the central corona may suffer significant absorption by these outflows. This scenario resembles a Seyfert 2 galaxy with some radio emission.

For $\eta$ between $0.7$ and $1.0$, $\theta < 33^\circ$, and blob energy larger than $0.4$, a jet may form.
A large $\eta$ implies a large ADAF, which pushes the thin disk outward; consequently, less UV flux and fewer broad lines are expected.
However, when a jet forms, additional X-ray and UV emission may arise from the jet base due to shocks. This emission could illuminate the BLR blobs, the thin disk, and possibly the ADAF disk surface, leading to the reappearance of broad lines.

It shall be noted again that the $\eta$ values listed in the previous paragraphs are only illustrative and not derived from calculations. The actual situation is highly complex and depends on the source. The emission mechanism of the broad lines and the energy source are also more intricate than the simple picture adopted here. A detailed comparison between the model and observations would require case‑by‑case simulations. In this work, we provide only an overall description.

The accretion rate onto the central black hole depends on the size of the ADAF disk. As the ADAF expands, the fractional contact area between the ADAF disk and the black hole decreases, while more material is carried away by outflows or jets. Consequently, the true accretion rate must decline. Such a decrease in the true accretion rate is directly observed in the neutron star system Her X-1, where, after the anomalous low state, the neutron star’s spin period becomes longer than expected due to the reduced accretion rate (\citealt{Coburn2000}; \citealt{vrtilek2001}). The anomalous low state of Her X-1 is interpreted as a large ADAF size persisting over a long period (\citealt{Xu2026b}). Although Her X-1 is a neutron star system, the same reasoning should apply to black hole systems. The theoretical accretion rate derived from the outer thin disk condition is unaffected by variations in the size of the inner ADAF, the latter only influences the actual accretion rate onto the central black hole.

With these basic principles established, interpreting the changing-look AGN (CL-AGN) phenomena within the variable ADAF disk scenario \citep{Xu2026b} becomes very straightforward. \citet{Ricci2023} provided an excellent summary of changing-look AGN in his review, and many of the observational facts discussed below are drawn directly from that work. The definitions of changing-obscuration and changing-state AGN follow \citet{Ricci2023}. Changing-obscuration AGN (CO-AGN) are primarily due to changes in the $N_H$ column density along the line of sight, whereas changing-state AGN (CS-AGN) are due to changes in the AGN accretion process.

The variation of the $\eta$ parameter or the size of the ADAF can occur at any $\eta$ value, thus, CS-AGN may arise whenever $\eta$ changes. In contrast, changes in the column density $N_H$ predominantly take place in the range $0.4 < \eta < 0.7$, where outflows can easily form while the source is observed along a suitable line of sight. Consequently, CO-AGN mostly occur within this $\eta$ interval. For $0.4 < \eta < 0.7$, two cases are possible: first, $\eta$ remains unchanged and only outflows may obscure the source; second, both $\eta$ and the outflows vary. The latter case represents a mixture of CS-AGN and CO-AGN.

Overall, CS-AGN and CO-AGN are governed by changes in the size of the ADAF (or $\eta$), they are simply two sides of the same coin — two accompanying effects of an evolving ADAF disk. When the ADAF expands to a certain size, with $\eta > 0.7$ and the central funnel deepens, radio jets may form. The reverse process can also occur, as the ADAF contracts from a large to a small size. Given that the ADAF is not stable, oscillations in its size may occur, rather than a unidirectional change.

The above is only a simple overall sketch of the relation between ADAF size and CL-AGN. We will discuss it in detail, source by source, in the next chapter.

\section{Comparison with observations}

\subsection{1ES 1927+654}

The changing-look AGN 1ES 1927+654 (CS-AGN) underwent a dramatic multi-wavelength outburst beginning in late 2017. Initially, the source exhibited a luminous UV/optical flare, reaching peak magnitudes approximately four magnitudes above its quiescent state \citep{Trakhtenbrot2019}. Following this UV peak, new and strong broad Balmer emission lines appeared in the optical spectrum, signaling a type transition from a Seyfert 2 to a Seyfert 1 \citep{Trakhtenbrot2019}. In a surprising turn, the X-ray emission, initially comparable to pre-outburst levels, then dropped precipitously by nearly four orders of magnitude over about 100 days \citep{Ricci2020}. This X-ray dip coincided with the near-complete disappearance of the power-law component, implying the destruction of the X-ray corona \citep{Ricci2020}. Approximately 300 days after the initial UV event, the X-ray flux began to recover, and the power-law component reappeared, indicating the re-formation of the corona \citep{Ricci2020}. Ultimately, the source faded back to its pre-outburst state, with the broad emission lines disappearing again, completing the return to its original Seyfert 2 classification.

The source underwent multi-wavelength observations from January 2022 to May 2023. The key finding is the re-emergence of a bright soft X-ray state, approximately four years after the initial event. The soft X-ray (0.3–2~keV) flux increased by a factor of $\sim$5 over $\sim$1~year, reaching a level comparable to that observed during the peak of the 2018 outburst. However, this re-brightening was not accompanied by corresponding changes in other bands: the UV flux increased by only $\lesssim30\%$ \citep{Ricci2020}.

The source also exhibited a dramatic late-time radio brightening beginning in February 2023, approximately five years after the initial optical/UV outburst \citep{Ricci2020}. The radio flux density increased by a factor of $\sim$60 compared to pre-outburst levels, reaching $\sim$30~mJy at 6~GHz \citep{Meyer2025}. High-resolution VLBA observations revealed, for the first time in this source, an extended bipolar radio jet spanning $\sim$0.15~pc on each side of the core. The source is viewed edge-on, with an inclination angle of about $85^\circ$ \citep{Gallo2013}.

Before the UV flare, the source is a Type 2 AGN with no radio emission. We may assume its $\eta \sim 0.4$. During the UV flare, $\eta$ decreases, and the ADAF also decreases, causing the thin disk to extend inward. As a result, the BBB component increases, and the UV emission rises. The increase in UV further induces the broad emission lines. The source thus turns into a Type 1 AGN in 2018. When the broad-line flux reaches its peak, the X-ray flux drops by four orders of magnitude and then recovers as the broad-line flux declines \citep{Trakhtenbrot2019}. \citet{Ricci2020} concluded that the X-ray corona disappears during the X-ray dip, but the mechanism they invoke, i.e., "debris stream colliding with an accretion disk", is unconvincing. In our variable ADAF model, when the ADAF shrinks, the accretion rate is expected to increase. If the corona remains stable (i.e., its size does not change) and the viewing angle is nearly edge-on, more of the corona becomes visible, and thus the X-ray flux should increase. However, the observed flux decrease instead supports the reduction or disappearance of the corona. The reason for the corona's decrease is as follows. When $\eta$ is small, say $0.3$, the blob energy from the funnel is $2\eta-1 = -0.4 < 0$. In this case, blobs cannot escape the black hole's gravity; instead, they collide with one another, forming a corona around the black hole that remains confined to the vicinity of the funnel. As $\eta$ decreases, both the ADAF and the corona shrink, leading to a drop in X-ray emission. The accretion flow then falls into the black hole very close to or within the ISCO, resulting in low radiation. When $\eta$ reaches $0$, the ADAF disappears, and a thin disk forms around the black hole, accreting material onto it. The corona's spectrum is primarily determined by the turbulence strength, i.e., the value of $\eta$ in the accretion flow. As discussed in \citep{Xu2026b,schlickeiser2002}, the turbulent flow establishes an energetic particle distribution as a power law with a typical index of $2$. When $\eta$ decreases, turbulence decreases, so the power-law component weakens, and the blackbody component becomes dominant during the deepest X-ray dip.

However, the thin disk is not stable, as the released binding energy is high, and the ADAF will form again. Then the X-ray corona will re-form, the thin disk will shrink outward, the UV emission will decrease, and the broad lines will also decline, exactly as observed (Fig. 2 in \citealt{Trakhtenbrot2019}). The disappearance and reappearance of the ADAF occur very often in X-ray binary systems \citep{Inoue2022,ingram2019}. The offset between the UV peak and the X-ray dip may be due to time lags.

During the UV/optical flaring phase, the radio flux decreases by a factor of about four, followed by a rebrightening \citep{Laha2022, Ricci2023}, corresponding to the phase in which the ADAF reduces and then increases.
The formation of radio jets in 2025 \citep{Meyer2025} suggests that the $\eta$ value exceeds 0.5, possibly reaching above 0.7, while the ADAF becomes larger. We propose that there existed a period when $\eta$ was just above 0.5; during such a phase, outflows would erupt, and the absorption by $N_H$ would increase — an effect that could be confirmed if X-ray data were available.

\subsection{Mrk 590}
Mrk~590 is a prototypical changing-look AGN (CS-AGN) that has undergone dramatic transitions in its spectral state over the past several decades. Historically classified as a classic Seyfert~1 galaxy, multi-wavelength observations spanning more than 40 years have revealed a remarkable decline in its nuclear activity \citep{Denney2014}. The continuum luminosity has decreased by a factor of $\sim$100 relative to its peak, and the once-prominent broad emission lines in the UV/optical spectrum have all but disappeared, leaving Mrk~590 currently classified as a Seyfert~1.9--2 with only a weak broad H$\alpha$ component \citep{Denney2014, Ricci2023}.

Following its near turn-off in 2012, Mrk~590 partially re-ignited in 2017, entering an unusual repeating flaring state \citep{Lawther2023}. The data demonstrate that Mrk~590's changing appearance is a consequence of a genuine decrease in the black hole accretion rate \citep{Denney2014}.

At radio wavelengths, Mrk~590 reveals a compact core with no significant extended jet-like features. A comparison of radio flux densities between the 1990s and 2015 shows a significant decline of tens of percent, correlating with the decline in optical-UV and X-ray luminosities over the same period. This correlated variability demonstrates the AGN accretion--outflow connection and confirms that the changing-look behavior originates from variable accretion rates rather than dust obscuration \citep{Koay2016}. The inclination angle of the accretion disk is $i \sim 38^\circ$ \citep{Lawther2025}.

According to the observational results from these works \citep{Denney2014,Lawther2023,Lawther2025,Mandal2021}, the $\eta$ value and the size of the ADAF disk in Mrk~590 underwent the following evolutionary phases. From 1973 to 1983, $0.2<\eta<0.4$, with the source in Type 1, exhibiting strong broad H$\alpha$ and relatively weak H$\beta$; from 1983 to 2000, $0.1<\eta<0.3$, in a more typical Type 1 state with UV emission and strong H$\beta$; from 2000 to 2017, $0.3<\eta<0.7$, in Type 2, showing little UV and no (or weak) broad lines; after 2017, $\eta$ decreased to possibly around 0.2, accompanied by the reappearance of broad lines. As discussed in \S2, the true accretion rate is inversely proportional to the $\eta$ value, and the same holds for the X-ray and UV fluxes. The situation in Mrk~590 differs from that in 1ES 1927+654 due to their different inclination angles: in Mrk~590, $i\sim 38^\circ$, so the corona and UV emission are viewed more directly, whereas in 1ES 1927+654, $i\sim 85^{\circ}$, the source is viewed nearly edge-on, and the corona may be obscured by the ADAF disk. The difference in viewing angle plays an important role in the observed properties. Moreover, in 1ES 1927+654, the $\eta$ value appears to have dropped to 0 over a short period, leading to the disappearance of the X-ray corona, a behavior not observed in Mrk~590. The radio emission from Mrk~590 is linked to the core activity but not to a jet (since no jet is present in Mrk~590), and is therefore generally related to the accretion rate. A detailed comparison with multi-wavelength observations at different epochs requires further comprehensive simulations based on the variable ADAF disk.

\subsection{NGC~4151}
NGC~4151 is a nearby Seyfert~1.5 galaxy (CO-AGN) that serves as a unique laboratory for studying the complex interplay between accretion disk emission, BLR dynamics, and variable X-ray absorption. Following the synthesis by \citet{Ricci2023}, the source exhibits dramatic X-ray absorption variability on short timescales \citep{Puccetti2007}. These variabilities are 10 to 100 times shorter than those typically observed in Seyfert~2 galaxies and imply that the obscuring clouds are located within $\sim$1 light-day of the central engine, consistent with the innermost broad-line region \citep{Puccetti2007}.

Subsequent multi-epoch observations during 2017–2018 measured neutral X-ray absorption column densities of $N_{\rm H} \sim 1.2$–$3.4 \times 10^{23}~{\rm cm}^{-2}$ \citep{Kumar2024}. These column densities are $\sim$100 times larger than those inferred from UV extinction assuming the Galactic dust-to-gas ratio, suggesting that the X-ray obscurer is more compact than the UV-absorbing material and may consist of dust-free clouds located within the sublimation radius \citep{Kumar2024}. The absence of a correlation between variations in X-ray absorption and UV emission further supports this compact geometry \citep{Kumar2024}.

Modeling of velocity-resolved variations through reverberation mapping of the H$\beta$ broad emission line reveals that the BLR is well described by a very thick disk with an opening angle $\theta_o \approx 57^{\circ}$ and an inclination angle $i \approx 58^{\circ}$, indicating that our line of sight skims just above the BLR surface \citep{Bentz2022}. The similarity between the inclination and opening angles is intriguing in light of previous observations suggesting that BLR gas temporarily eclipses the X-ray source.

High-resolution Very Long Baseline Array (VLBA) observations have identified a compact, flat-spectrum $\sim$3~mJy radio source located at the position of the AGN and its central black hole. This radio core has an upper size approximately ten times the diameter of the broad-line region \citep{Ulvestad2005}. Despite being radio-quiet, NGC~4151 hosts a remarkable two-sided sub-parsec radio jet. The VLBA imaging reveals a 0.2~pc two-sided base to the well-known arcsecond-scale jet, with the jet initially propagating north-northeast before turning sharply eastward. The apparent speeds of jet components relative to the radio core are $<0.050c$ and $<0.028c$ \citep{Ulvestad2005}. Follow-up high-resolution e-MERLIN observations have revealed intriguing variability in the jet structure. While the core component (C4W) remained constant between epochs within uncertainties, the easternmost component increased in peak flux density from $19.35 \pm 1.10$ to $37.09 \pm 1.86$~mJy~beam$^{-1}$ \citep{Williams2020}.

Given the detection of weak radio twin-jets and weak broad lines, we can constrain the $\eta$ value to lie between 0.4 and 0.8. It is evident that outflows may play an important role in this source. An $\eta = 0.7$ corresponds to a funnel opening angle of $33^{\circ}$, which is equivalent to an angle of $57^{\circ}$ with respect to the accretion disk, i.e., $\theta_o \approx 57^{\circ}$. Since the inclination angle is $i \sim 58^{\circ}$ \citep{Bentz2022}, nearly equal to $\theta_o$, the outflow is expected to significantly obscure the central source, frequently causing large-amplitude variations in the $N_H$ column density and X-ray flux. The outflow may block the central source and parts of the broad-line region (BLR), but not the UV-emitting regions (i.e., the big blue bump region in the thin disk); therefore, the UV flux does not vary in tandem with the X-ray flux.

\subsection{NGC 1365}
NGC~1365 is an archetypal obscured AGN (Type 1.8, CO-AGN) that exhibits dramatic X-ray absorption variability on short timescales, driven by clouds in the broad-line region (BLR) crossing the line of sight to the central engine \citep{Risaliti2009}. Based on the pivotal synthesis by \citet{Ricci2023}, the source displays rapid changes in column density on timescales of hours to days, with values spanning N$_H$ $\sim$ 10$^{22}$ to $10^{24}$ cm$^{-2}$, transitioning between Compton-thin and Compton-thick states. This variability is not due to intrinsic luminosity changes but rather to occultation by compact, clumpy obscurers located at distances $R \sim 10^{16}$ cm from the black hole, consistent with the BLR scale \citep{Brenneman2013}. The absorbers exhibit cometary morphologies with dense heads and ionized tails, suggesting a dynamic, outflowing medium \citep{Maiolino2010}. The near-constant UV emission, in contrast to the highly variable X-ray absorption, indicates that the obscuring clouds are compact and preferentially cover only the compact X-ray corona rather than the more extended UV-emitting accretion disk \citep{Swain2023}. This behavior establishes NGC~1365 as a key laboratory for studying the clumpy circumnuclear gas and disk-wind connections in active galaxies.

As discussed in \S2, type~2 AGN requires $\eta > 0.2$, no jet formation expects $\eta < 0.7$, and a source with major outflows from the central funnel needs $\eta > 0.5$. Thus, the observational facts constrain the $\eta$ value of NGC~1365 to lie between 0.4 and 0.7. The inclination angle of the accretion disk is approximately $i \sim 60^\circ$ \citep{Walton2014}. For $\eta \sim 0.6$, the funnel angle is about $\sim 40^\circ$, or $50^\circ$ above the disk plane. Hence, the outflows are about $20^\circ$ above the line of sight. It is very likely that the outflows can block the line of sight toward the corona, making this source a typical CO-AGN. The cometary shape of the blocking clouds is not surprising, as they are outflows. These outflows have two capabilities: they can serve as BLR clouds and move as clouds to block the X-ray corona, as suggested in \citep{Ricci2023}.

\subsection{3C~390.3}
3C 390.3 is a well-studied broad-line radio galaxy (BLRG) that has exhibited long-term spectral variability in both the optical and X-ray bands. As one of the few candidate radio-loud changing-look AGN (belongs to CS-AGN type), it offers an important contrast to the predominantly radio-quiet population. According to the synthesis by \citet{Ricci2023}, this source displays remarkable multi-wavelength properties that help bridge our understanding of radio-quiet and radio-loud active galactic nuclei.

The optical spectrum of 3C~390.3 is characterized by double-peaked Balmer emission lines (H$\alpha$, H$\beta$, H$\gamma$), which provide direct evidence for gravitationally bound orbital motion of the broad-line region gas \citep{Sergeev2017}.
In the X-ray band, BeppoSAX observations detected, for the first time, both the 6.4 keV iron K$\alpha$ line and a strong reflection hump, produced by illumination of cold material by the primary X-ray continuum. A historical study reveals variable cold absorption along the line of sight, with changes in column density $N_{\mathrm{H}}$ that are not correlated with continuum variations, suggesting geometric changes in the absorber rather than variations in its ionization state \citep{Grandi1999}.

High-resolution VLBI observations at 5~GHz reveal a compact radio core with a core-jet structure, aligning with the 300~kpc double-lobe structure \citep{Preuss1980}. The jet exhibits subluminal apparent motions $<0.5c$, and the inclination angle $i$ of the accretion disk is approximately $26^{\circ}$ \citep{Eracleous1996, Preuss1980}. 

3C~390.3 differs from the other sources listed above in that it is a radio-loud CS-AGN. The parameter $\eta$ is expected to be larger than 0.7. Typically, for large $\eta$, the thin disk is pushed outward, so less UV and broad-line emission is expected. An alternative is that the UV emission arises from jet activity. This view is supported by the direct link between the variable optical continuum and the subparsec-scale jet in 3C~390.3 \citep{Arshakian2010}. The funnel angle for $\eta > 0.7$ is less than $33^{\circ}$, while the inclination angle $i$ is about $26^{\circ}$; therefore, variable $N_{\rm H}$ column density is expected to be observed frequently in this source, which may obscure the X-ray corona.

\subsection{GSN~069}
GSN~069 is an X-ray quasi-periodic eruption (QPE) source discovered by \citet{Miniutti2019}. This source exhibits quasi-periodic soft X-ray eruptions with a recurrence time of $\sim 9$ hours and a duration of $\sim 1$ hour. During an eruption, the X-ray count rate increases by up to two orders of magnitude, and the X-ray spectrum oscillates between a cold phase (about 50 eV) and a warm phase (about 120 eV). The mass of the central black hole is approximately $4\times10^5\,M_{\odot}$ \citep{Miniutti2019}, and the inclination angle of the source is constrained to $30^{\circ} < i < 65^{\circ}$ \citep{Guolo2025}.

The QPE phenomenon can be understood within the framework of a variable ADAF. The characteristic temperature of a thin accretion disk is given by $T(r) \approx \left( 3GM\dot{M} / 8\pi \sigma_{\text{SB}} r^3 \right)^{1/4}$, which scales as $T(r) \propto M^{-1/4}$ at a scaled radius $r\propto M$ with the assumption of $\dot{M}\propto M$, or $T(r) \propto r^{-3/4}$, where $M$ is the black hole mass and $\sigma_{\text{SB}}$ is the Stefan–Boltzmann constant \citep{Reynolds2003,Shakura1973}. At a radius of about $50 R_g$, where the ADAF connects to the thin disk, the typical temperature is approximately $10^5$ K for $10^8\,M_{\odot}$ AGN, $10^6$ K for $10^4$–$10^5\,M_{\odot}$ AGN, and $10^7$ K for X-ray binaries. The corresponding spectral components are the UV/optical (BBB), UV/soft X-ray, and soft X-ray, respectively. The X-ray eruptions in GSN~069 are interpreted as the shrinking of the ADAF (or an inward extension of the thin disk). This naturally explains why the eruption spectrum is dominated by soft X-rays (120 eV), while the UV/X-ray (50 eV) component remains nearly unchanged \citep{Miniutti2019}.  In fact, under the assumption of multicolor blackbody emission from the thin disk, Fig. 2a and Fig. 3 in \citet{Miniutti2019} directly demonstrate the inward movement and subsequent recession of the thin disk during the eruptions.  After an eruption, the ADAF returns back to the original place and waits for next shrinking. Quasi-periodic variability has also been observed in the black hole X-ray binary GRS~1915+105 \citep{Belloni1997} with a shorter recurrence period, and in neutron star binary Hercules~X-1 \citep{vrtilek2001} with a much longer period of 35 days. The latter is explained as originating from a variable ADAF in \citet{Xu2026b}.

The parameter $\eta$ of GSN~069 is expected to lie in the range of 0.1–0.4, suggesting that this source should be classified as a Type I (or mixed Type II) AGN. The absence of observed broad emission lines \citep{Miniutti2019} can be attributed to the fact that the illuminated UV region is located approximately ten times farther (scaled with $R_g$) from the central source (as compared with more massive AGN), resulting in a weaker emission.

\subsection{Sources With Interesting Features}
We discuss several other interesting sources below, in addition to the classic ones listed above.

\textbf{SDSS~J101152.98+544206.4 (CS-AGN)} underwent a Type 1 $\rightarrow$ Type 1.9 $\rightarrow$ Type 1 cycle from 2002 to 2024 (via 2012), within a timescale of about 20 years \citep{Lyu2025}. Its $\eta$ value evolved through 0.2--0.4--0.3, while the ADAF size went from small to large to intermediate. The optical emission correspondingly passed through high, low, and intermediate stages, tracking the size of the thin disk and the accretion rate. The accretion rate varied opposite to $\eta$ (thus high--low--intermediate), and the infrared followed the accretion rate.

\textbf{Mrk~1018 (CS-AGN)} underwent a Type 1.9 $\rightarrow$ Type 1 $\rightarrow$ Type 1.9 cycle from 1979 to 2016 (via 1985--2015), within a timescale of nearly 40 years \citep{McElroy2016,Ricci2023}, showing an opposite trend compared to SDSS~J101152.98+544206.4. Its $\eta$ value evolved through 0.4--0.2--0.4. The X-ray, UV, optical continuum, and broad-line fluxes are correlated, while its inclination angle $i \sim 60^{\circ}$ \citep{Noda2018}. This source is very similar to Mrk~590.  The radio luminosity scales with the X-ray luminosity during the transitions, suggesting a common origin in the accretion disk corona, and is linked to the true accretion rate determined by the $\eta$ value.

\textbf{NGC~1566 (CS-AGN)} is a nearby Seyfert~1 galaxy that has exhibited repeated, cyclic changes in its AGN luminosity over decadal timescales (four cycles between 1970 and 1985) \citep{Alloin1986}. During these transitions, the narrow-line flux remained constant, while the broad lines and continuum varied. A burst has a typical rise time of 20~days, followed by a longer exponential decay of approximately 400~days. This fast rise and slow decay resemble the behavior observed in X-ray binaries \citep{Inoue2022,Xu2026b}.
A recent changing-state from type~1.9–1.8 to type~1.2 was observed in 2018 \citep{Oknyansky2019,Parker2019}. Multi-wavelength observations show that the X-ray, UV, optical, and broad-line fluxes vary proportionally. The inclination angle of NGC~1566 is $i \leq 11^\circ$, making it nearly face-on.
The cyclic changing-state of this source is linked to variations in the $\eta$ parameter within a range of 0.4–0.2, where the burst phase and broad lines correspond to the smaller $\eta$ value. This source is similar to Mrk~1018 and Mrk~590 in many ways, except their inclination angles are quite different. Although the source is nearly face-on and outflows are typically weak, they may still be detectable under certain conditions \citep{Parker2019}.

\textbf{PG~1211+143 (CS-AGN)} is a prototypical narrow-line Seyfert 1 galaxy (H$\beta$ FWHM < 2000 km/s) that exhibits one of the most compelling examples of an ultra-fast outflow (UFO) from an active galactic nucleus \citep{King2015}. Early XMM-Newton observations revealed deep blue-shifted absorption lines in its X-ray spectrum, primarily associated with highly ionized iron, providing the first clear evidence for a high-velocity ionized outflow in a non-broad absorption line AGN \citep{Pounds2003,King2015}. The outflow is characterized by a relativistic velocity of $v \sim 0.1$--$0.15c$ and a high ionization state, with the dominant absorption feature identified as arising from He-like Fe~XXV (rest-frame energy $\sim 6.7$ keV) \citep{King2015}.
The parameter $\eta$ is expected to be in the range of 0.4--0.7, combined with the relatively narrow H$\beta$ line and fast outflows. Thus the opening angle should be within $33^{\circ}$ to $51^{\circ}$. This range corresponds to Type II AGN, but the presence of narrower broad lines is not surprising. The inclination angle is estimated to be $\sim 28^{\circ}$ \citep{Zoghbi2015}, so this source can be viewed with outflows near or just below the line of sight. The blue-shifted Fe absorption lines are likely due to the random outflows moving toward the observer from wide-opening-angle funnels.

\section{PS/GPS and CSS radio sources}

Peaked-spectrum (PS) sources, along with their subclass Gigahertz peaked-spectrum (GPS) sources, and compact steep-spectrum (CSS) radio sources constitute a category of compact, powerful radio objects that have both challenged and advanced our understanding of active galactic nucleus (AGN) evolution over the past three decades. As comprehensively reviewed by \citet{Odea1998} and \citet{Odea2021}, these sources are characterized by their distinct radio spectral shapes and compact physical sizes, placing them at a critical evolutionary stage in the lifecycle of powerful radio galaxies. PS and CSS sources correspond to intermediate-to-large ADAF disk sources, with the typical parameter $\eta$ lying in the range $0.6 < \eta < 0.8$, and with $\eta$ of CSS being larger than that of PS. By comparison, radio-loud AGNs with extended jets tend to have $\eta$ values exceeding $0.8$. Since the jet speed is proportional to $\eta$, it is expected that jet speeds in PS and CSS sources are lower than those in radio-loud AGNs.

The defining characteristic of these sources is a convex (peaked) radio spectrum resulting from synchrotron self-absorption, though free-free absorption through an inhomogeneous screen may also contribute \citep{Odea1998}. The turnover frequency $\nu_m$ scales with the linear source size $l$ as $\nu_m \propto l^{-0.65}$, indicating a fundamental physical relationship between these parameters \citep{Odea1998}.

The classification distinguishes two subpopulations based on size and spectral peak frequency. GPS sources peak near 1~GHz (observer frame) and are entirely contained within the narrow-line region, with sizes $\lesssim 1$~kpc. CSS sources peak at lower frequencies ($\lesssim 500$~MHz) and are somewhat larger (1--15~kpc), though still confined within the host galaxy \citep{Odea1998}. These sources constitute significant fractions of the bright radio source population: approximately 10\% are GPS sources and 30\% are CSS sources \citep{Odea1998}.

The radio morphologies of GPS and CSS sources closely resemble those of large-scale classical doubles, exhibiting compact symmetric structures. Nevertheless, some sources display distorted morphologies, indicating possible interactions with their ambient environments \citep{Odea1998}. Their low polarization and high rotation measures are consistent with these sources being embedded in a dense interstellar medium \citep{Odea2021}. Such dense environments may be related to outflows from the central engine, which are expected to be strong for $\eta \sim 0.6-0.8$.

At infrared wavelengths, the observed properties are consistent with stellar populations and AGN bolometric luminosities comparable to those of 3CR classical doubles. In CSS sources, high surface brightness optical emission (dominated by emission-line gas) is aligned with the radio axis at all redshifts, providing evidence for jet–cloud interactions \citep{Odea1998}. The optical emission-line properties also indicate the presence of dust in the emission-line regions. X-ray observations of high-redshift GPS quasars and some GPS galaxies reveal significant gas columns toward the nuclei \citep{Odea1998}. All these findings are consistent with a scenario of enhanced outflows around $\eta\sim 0.6-0.8$.

The nature of GPS and CSS sources has been debated since their discovery. \citet{Odea1998} articulated two competing models: (1) they are ``frustrated'' sources, perpetually confined by interaction with dense gas in their environments, or (2) they are young sources that will evolve into large-scale radio galaxies. It is not easy to constrain the age of GPS and CSS sources based on the ADAF size. However, it is true that the size of the ADAF may vary, potentially causing the source to transition between different types of AGNs. Variations in ADAF size are more readily observed in X-ray binary systems, where the black hole mass is smaller and the evolution proceeds more rapidly (compared to the human lifetime) \citep{Xu2026b}.

Evidence for episodic AGN activity over a wide range of timescales has also been identified in some GPS/CSS sources, suggesting that the duty cycle of AGN activity may be an important factor in determining the observed source properties \citep{Odea2021}. Within the context of the variable ADAF model, and given that only $\eta > 0.7$ may produce significant jets, such episodic AGN activity is not surprising.

As discussed earlier, for large $\eta$, PS/CSS sources are not expected to exhibit broad emission lines. Nevertheless, if the jets produce strong UV/X-ray radiation, broad emission lines may still be observed, as seen in the radio-loud AGN 3C~390.3. Among GPS sources, Mrk~668 provides an example showing a broad H$\alpha$ emission line \citep{Eracleous1994}.

Overall, PS and CSS radio sources are associated with an ADAF disk, with typical $\eta$ values likely in the range of 0.6–0.8. The observed interactions between the radio source and the emission-line gas, the presence of dust in the emission-line regions, and the evidence of significant gas columns toward the nuclei in these sources \citep{Odea1998} are all consistent with an intermediate-to-large ADAF disk scenario.

\section{Summary and Discussions}

The variable ADAF disk model successfully accounts for a range of observational phenomena in X-ray binary systems~\citep{Xu2026a} and appears capable of explaining most CL-AGN phenomena as well.
Owing to the large difference in central black hole mass, XRBs evolve more rapidly than AGNs. Compared to black hole XRBs where cycles of X-ray bursts and decays typically span the full range of ADAF size variations, CL-AGNs exhibit a more limited range of ADAF variability.

We use the parameter $\eta$ introduced in \citep{Xu2026a} to discuss, in a semi-quantitative manner, the variability of the ADAF disk size, changes in the $N_H$ column density, obscuration by outflows, correlated fluxes between X-ray, UV, and optical continua, the turn-on and turn-off of broad emission lines, radio jet formation, and the true accretion rate for different types of change-look AGNs. We find that Type~1, Type~2, and radio-loud AGNs are fundamentally governed by the $\eta$ parameter, ranging from small ($\sim 0.1$) to intermediate ($\sim0.4$) to larger ($\sim 0.9$) values.

In this framework, CO-AGNs and CS-AGNs represent two manifestations of the same CL-AGN phenomenon, driven by variability in the ADAF size (or the $\eta$ parameter). CO-AGNs occur mostly in the range $0.4<\eta<0.7$, where outflows are most frequent, whereas CS-AGNs can occur at any $\eta$ value provided that some instability takes place. We note that \citet{Sniegowska2020} proposed a toy model similar to ours, using the classic ADAF from \citet{Narayan1994} for the repeating CS-AGNs NGC~1566, NGC~4151, and NGC~5548. The ADAF in \citet{Narayan1994} is optically thin, whereas the one in this work is optically thick. There are also many other differences between these two ADAFs \citep{Xu2026a,Xu2026b}.

We did not discuss in detail the role of the central black hole mass in this work (but see discussion on GSN~069), but it surely plays an important role in the evolution of CL-AGNs. The more/less massive the central black hole, the larger/smaller the inner radius of the disk, and the lower/higher the typical disk temperature. The peak wavelength of the big blue bump may shift to longer/shorter wavelength, which will certainly change some behaviors of CL-AGNs.

PS and CSS radio sources are considered as young AGNs or transient sources~\citep{Odea1998}, and are regarded as AGNs with $\eta$ values around 0.6–0.8 in this work.

Numerous works have been devoted to the unification model of various AGN types (including blazars, which we did not discuss here) based on orientation effects~\citep{Antonucci1993,Urry1995}. It is evident that the size of the ADAF disk (or the $\eta$ parameter) also plays an important role in the unification scenario.

Although the variable ADAF disk model appears successful in explaining phenomena in both X-ray binaries and changing-look AGNs, the trigger for ADAF variability remains poorly understood. In this work, only semi-analytical and semi-quantitative analyses have been discussed. Further theoretical and observational studies are urgently needed.

\section*{Acknowledgements}

This work is supported by the China Manned Space Program Grant No. CMS-CSST-2025-A19 and Grants allocated for the development of the Multi-Channel Imager and the Integral Field Spectrograph, led by Drs. Zhenya Zheng and Lei Hao respectively, for the Chinese Space Station Telescope (CSST).

\section*{Data Availability}

No new data were created or analysed in this study.




\bibliographystyle{mnras}
\bibliography{example} 





\bsp	
\label{lastpage}
\end{document}